# Correcting Emittance Growth Due to Stray Sextupole Fields[1]
David H. Dowell
Stanford Linear Accelerator Center
Feb. 2018

**Abstract**
This paper discusses the emittance growth produced by sextupole fields in conjunction with a solenoidal focusing. The sextupole emittance is derived and compared with numerical simulations using a measured sextupole field for a vacuum chamber in a solenoidal field. It is found that the sextupole emittance is dominated by the field's radial non-linearity and the growth due to the skew angle is much smaller. It is shown that a corrector sextupole placed after the sextupole-solenoid system is effective at cancelling this emittance growth. Leaving only the growth due to the solenoid's spherical aberration.

**Derivation of the sextupole emittance**
The Lorentz force equation is
$$\vec{F} = e\vec{E} + e\vec{v} \times \vec{B} \tag{1}$$
If there is no electric field, then a magnetic field produces the following force in the x-direction
$$F_x = \frac{dp_x}{dt} = e(v_y B_z - v_z B_y) \tag{2}$$
Assume the fringe field effects are negligible and there is only the sextupole field. Then $B_z = 0$, and the first term becomes zero. In addition, we assume the paraxial approximation: $v_x, v_y \ll v_z = \beta c$, which allows writing
$$dz = \beta c\, dt \tag{3}$$
and converting the force equation into the following ray equation,
$$\frac{dp_x}{dz} = -eB_y \tag{4}$$
H. Wiedemann [1] gives the y-component of the sextupole field as
$$B_y(x,y) = \frac{1}{2} \left.\frac{\partial^2 B_y}{\partial x^2}\right|_{x,y=0} (x^2 - y^2) \tag{5}$$
The second partial of $B_y$ with respect to $x$ is to be evaluated on the z-axis which is also the beam axis. The factor of ½ assures that
$$\frac{\partial^2}{\partial x^2} B_y(x,y) = \left.\frac{\partial^2 B_y}{\partial x^2}\right|_{x,y=0} \tag{6}$$
The transverse momentum kick given to the beam by the sextupole is then the integral of the ray equation
$$\int_{p_x}^{p_x+\Delta p_x} dp_x = \Delta p_x = -\frac{e}{2} \int_{-\infty}^{\infty} \left.\frac{\partial^2 B_y}{\partial x^2}\right|_{x,y=0} (x^2 - y^2)\, dz \tag{7}$$

$$\Delta p_x(x,y) = -\frac{e}{2} \left.\frac{\partial^2 B_y}{\partial x^2}\right|_{x,y=0} L_{eff}\, (x^2 - y^2) \tag{8}$$
Here the sextupole field gradient, $\left.\frac{\partial^2 B_y}{\partial x^2}\right|_{x,y=0}$, is assumed to be constant over an effective length of $L_{eff}$ and zero everywhere else.

The normalized emittance for an upright phase space ellipse is
$$\epsilon_n = \frac{\sigma_x \sigma_{p_x}}{mc} \tag{9}$$
$\sigma_x$ is the rms size of the beam in the x-direction and $\sigma_{p_x}$ is the rms x-momentum of the beam. The sextupole emittance is found by computing the transverse rms momentum from the variance of $\Delta p_x$, and assuming a Gaussian transverse distribution with an rms of $\sigma_x$. The variance of the x-momentum is then
$$\sigma_{p_x}^2 = \frac{\int_{-\infty}^{\infty} [\Delta p_x(x,y=0)]^2 e^{-\frac{x^2}{2\sigma_x^2}} dx}{\int_{-\infty}^{\infty} e^{-\frac{x^2}{2\sigma_x^2}} dx} \tag{10}$$
Computing the integrals, taking the square root, and multiplying by the rms beam size gives the emittance due to a sextupole field for beam with a Gaussian transverse distribution,

---

[1] A previous version of this paper has been published as SLAC Report SLAC-R-1091.

$$\epsilon_{n,sextupole} = \frac{\sqrt{3}}{2}\sigma_x^3 \frac{e}{mc} L_{eff} \left.\frac{\partial^2 B_y}{\partial x^2}\right|_{x,y=0} \tag{11}$$

Here $\sigma_x$ is the rms beam size at the sextupole field and $L_{eff}$ is the effective length of the sextupole field.

Thus, the normalized emittance grows as the beam size cubed. It can be distinguished from the spherical aberration emittance which scales like the beam size to the 4$^{th}$ power. This sextupole emittance does not include transverse coupling due to the beam's rotation in the solenoid or if the sextupole is skewed. It is shown later that the skew of a sextupole field has a minor effect on the emittance growth of a sextupole+solenoid system. However, it is important to include the skew angle when cancelling the emittance growth with corrector sextupole.

**Comparison of Theory and Simulation Using a Measured Sextupole Field**

The SLAC magnetic measurements group used a rotating coil to measure the sextupole field of a solenoid lens with its beamline vacuum pipe installed. A significant sextuple field was measured, and it was determined that most of the sextupole field was from regions with high magnetic permeability in the walls of the vacuum pipe. These magnetic patches concentrate the solenoid's magnetic flux lines to locally perturb the solenoid's magnetic field and produce a sextupole field pattern at the beam axis. In general, a patchwork of permeabilities can and will produce other magnetic multipoles. Therefore, it is important to study how each affects the emittance. This paper discusses the sextupole multipole field. The quadrupole multipole has been discussed in a recent publication [2]. Future work will study the octupole and its effect on the beam emittance.

A long rotating coil was used to measure the multipole fields of the LCLS-II solenoid at a current of 10 amperes. (The nominal operating current of the solenoid is 8 amperes for an 800 KeV beam energy.) The rotating coil radius, $r_{coil}$, was 1 cm. The integrated second-order gradient of the magnetic field at the coil radius is given by

$$L_{eff}\left.\frac{\partial^2 B_y}{\partial x^2}\right|_{x,y=0} = \frac{2}{r_{coil}^2} \int B_3(I_{sol}, r_{coil})\, dz \tag{12}$$

Here $B_3$ is the sextupole field at the rotating coil radius, $r_{coil}$, and solenoid current, $I_{sol}$. The long coil measures the integral of this field over the length of the coil which is much longer than the field's length. Therefore, the long coil gives the total integrated sextupole field at the coil radius, which was found to be [3]

$$\int B_3(I_{sol} = 10\ amps, r_{coil} = 1\ cm)\, dz = 5.47 \times 10^{-7} Tm \tag{13}$$

And

$$L_{eff}\left.\frac{\partial^2 B_y}{\partial x^2}\right|_{x,y=0} = \frac{2}{(0.01m)^2} 5.47 \times 10^{-7} Tm = 0.0109\ T/m \quad \text{at } I_{sol} = 10 \text{ amps} \tag{14}$$

$$L_{eff}\left.\frac{\partial^2 B_y}{\partial x^2}\right|_{x,y=0}(I_{sol} = 10A) = 0.0109\ T/m = 109\ G/m \tag{15}$$

The simulation assumes the sextupole is 10 cm long with a sextupole field gradient of 0.109 T/m$^2$ to give the measured integrated sextupole gradient of 0.0109 T/m. Therefore

$$\left.\frac{\partial^2 B_y}{\partial x^2}\right|_{x,y=0}(I_{sol} = 10A) = \frac{109\ G/m}{L_{eff}=0.1m} = 1090\ G/m^2 = 0.109\ T/m^2 \tag{16}$$

When these parameters were used in a comparison of theory with numerical simulations, it was discovered that the GPT simulation program gave twice the emittance of Eqn. (11). Further investigation showed the factor of 2 comes from GPT's definition of the sextupole field. The GPT User Manual [4] defines the sextupole field without the ½ shown in Eqn. (5). The result is the GPT sextupole emittance is twice what it should be for the correct field. The correct sextupole field is defined in Eqn. (5). Therefore, the GPT sextupole field will be set to ½ times the field used in the theory to compensate for this difference in field definitions. For example, when Eqn. (11) uses a sextupole field gradient of 0.109 $T/m^2$ then GPT will be set to half that or 0.0545 $T/m^2$. The remainder of this paper will only quote the actual field and not the GPT-corrected field.

Figure 1 compares the sextupole emittance growth computed using Eqn. (11) with GPT simulations. There is good agreement between them with the above discussed factor of ½ multiplying the GPT sextupole field.

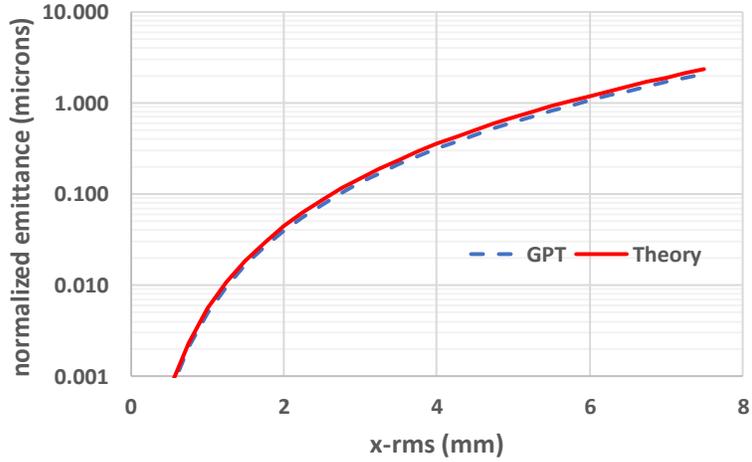

Figure 1: Comparison of emittances computed by Eqn. (11) (solid-line) with those simulated by GPT (dash-line). The sextupole field gradient is $0.109\ T/m^2$ and $L_{eff} = 0.10\ m$.

**Emittance Growth of a Rotated Sextupole Field**

Figure 2 plots the emittance growth of a single sextupole vs. the sextupole skew angle and shows a small modulation with a peak-to-peak amplitude of 0.023 microns (23 nm) repeating itself every 60 degrees as expected for this multipole. Thus, the coupling of the transverse dynamics due to skew is small for a sextupole, unlike that for a quadrupole where the coupling effects dominate. Instead the sextupole emittance growth is determined by the non-linear radial field. This means that the beam's rotation in the solenoid will generate minimal emittance growth although the skew angle introduced by the solenoid can be large. Again, this is very different behavior compared to the quadrupole fields where the modulation depth of the emittance would be 100%.

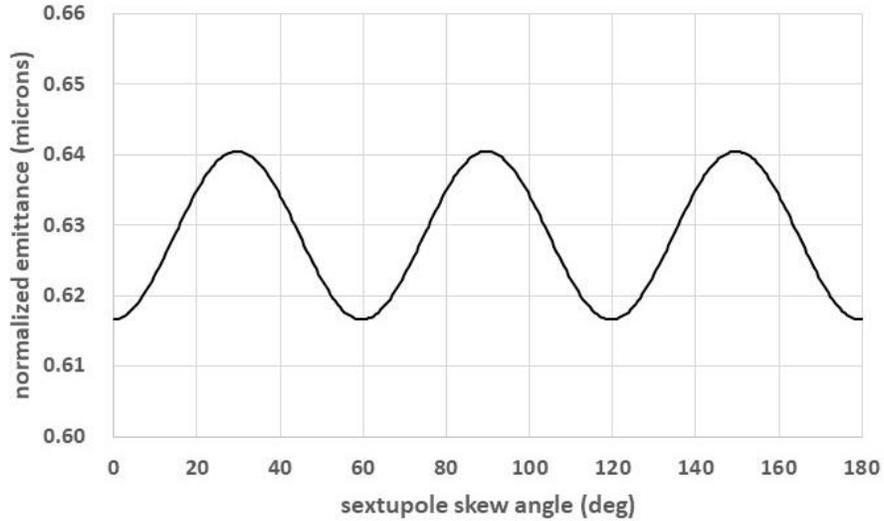

Figure 2: Sextupole emittance for $\sigma_x = 5$ mm as function of the skew angle. The beams size at the entrance of the sextupole is 5 mm-rms and the sextupole field parameters are the same as in Figure 1.

**Emittance growth of a sextupole + solenoid configuration**

The LCLS-I/GTF solenoid field map [5] was used in GPT with the solenoid field reduced in proportion with the lower beam energy of LCLS-II. The integrated solenoid field was set to 0.006187 T-m which focuses an 800 KeV beam approximately 90 cm after the solenoid. The solenoid effective length is 19.35 cm. The configuration being studied is shown in Figure 3.

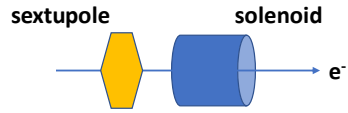

Figure 3: The sextupole+solenoid configuration simulated in GPT.

Figure 4 shows the simulation of the emittance growth vs. the sextupole rotation angle. Except for a ~10-degree phase shift and modulation variation, the sextupole emittance with the solenoid on is nearly the same as with the solenoid off. This again indicates that the growth is dominantly due to the field's radial non-linearity compared to the sextupole skew angle.

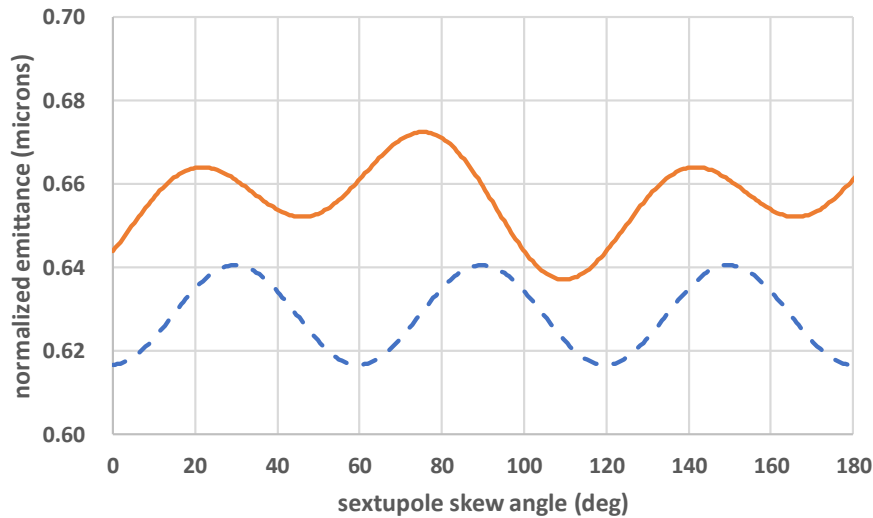

Figure 4: Simulated emittance for the sextupole+solenoid configuration (see Figure 3) vs. the sextupole skew angle. The solid-line is with the solenoid on and the dashed-line is with the solenoid off. The sextupole parameters are the same as those used in Figures 1 and 2.

**Cancellation of Sextupole Field Emittance Growth**

Figure 5 shows the sextupole-solenoid-sextupole configuration used to study the cancellation of a stray sextupole field and solenoid emittance growth with a corrector sextupole. In the simulations, sextupole1 produces the stray sextupole field. The solenoid then rotates this distorted phase space. This is followed by sextupole2 which will be the corrector sextupole. It is shown that a solution of the corrector sextupole's field strength and rotation angle can be found which completely cancels the emittance growth of a weak, stray sextupole field followed by a solenoid.

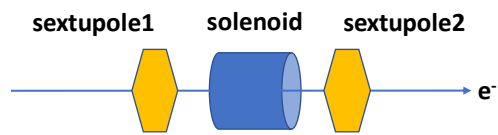

Figure 5: The sextupole+solenoid+sextupole configuration

The results of the emittance simulations are summarized in Figure 6.  In all cases the initial x-rms beam size is fixed at 5 mm, and sextupole1 is modeled with a field gradient 0.109 T/m$^2$ over an effective length of 0.10 m to duplicate the measured integrated gradient of 0.0109 T/m.  Short rotating coil measurements [3] show the sextupole field peaks near the ends of the solenoid with effective lengths of approximately 10 cm.

The solenoid field used in the simulations is the standard LCLS-1/GTF field map [5] with an integrated solenoidal field of 0.006187 T-m and an effective length of 19.35 cm.  Since the LCLS-I solenoid effective field length is longer than the LCLS-II solenoid's effective length, these simulations should be re-done using the measured LCLS-II field map.   However, it is expected that the results and conclusions presented here will be the same.  The difference in the field maps will become more important when the spherical aberrations of the solenoids are being studied.

Figure 6 shows the emittance simulations for the sextupole-solenoid-sextupole configuration in different modes of operation.  The nearly constant red- and dashed-curves are the emittance vs. sextupole1 skew angle with sextupole2 off and the solenoid off and on, respectively.  This small oscillation can be seen more clearly in Figure 4.  The yellow curve is for sextupole1 and sextupole2 both off and only the solenoid on.  This solenoid emittance growth is 0.23 microns and is due to the solenoid's 3$^{rd}$-order (spherical) aberration.

The grey, blue and green curves in Figure 6 show the emittance vs. sextupole2 skew angle for sextupole1 at an integrated field of 0.0109 $T/m^2$, and the solenoid at 0.006187 T-m. The three curves correspond to increasing field strengths of the sextupole2 corrector.  The grey, blue and green curves in the plot correspond to 1-times, 2-times, and 3-times the sextupole1 field, respectively.  The corrector field needs to be 3-times that of the stray field to cancel the emittance growth.  This is because the beam is smaller after the solenoid where the corrector sextupole is located and the emittance cancellation scales as the 3$^{rd}$ power of the beam size.

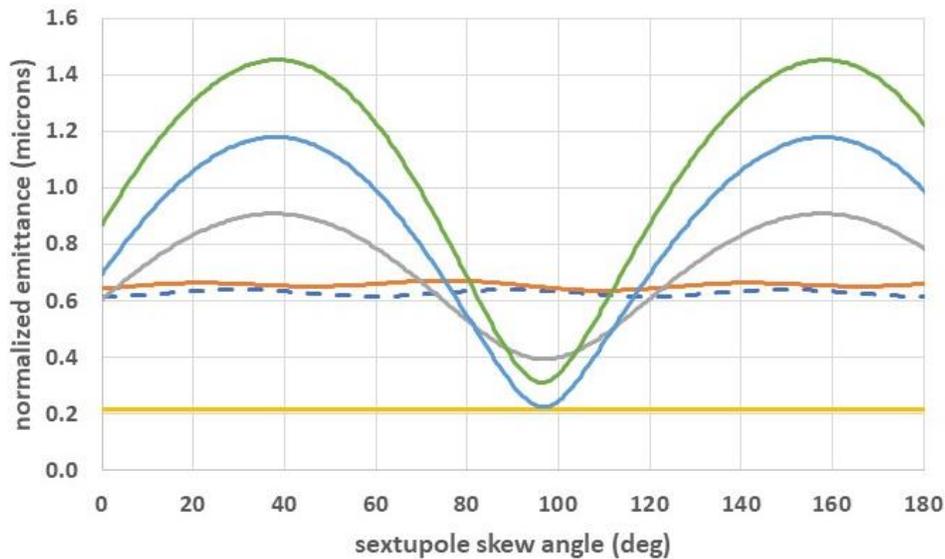

Figure 6:  The emittance growth vs. skew angle for the sextupole+solenoid+sextupole layout shown in Figure 5.  The initial x-rms beam size is 5 mm for all curves.  The red-curve is the emittance vs. the sextupole1 skew angle with only sextupole1 on.  The dashed-line is with sextupole1 and the solenoid on.  The grey, blue and green lines give the emittance growth for both sextupoles and the solenoid on.  The emittance is plotted vs. the sextupole2 rotation angle and correspond to the sextupole2 field being equal to 1-times, 2-times, and 3-times the sextupole1 field, respectively.  The yellow-line is the emittance with both sextupoles off and only the solenoid on.  This emittance results from the spherical aberration of the solenoid.

**Summary and Conclusions**

This paper investegated the emittance growth due to stray sextupole fields in conjunction with solenoidal focusing.   An expression for the sextupole emittance growth was derived and shown to compare well with simulations. During this work, it was found that the GPT program defines a sextupole field which is twice the accepted definition. To correct this issue, the GPT sextupole field in the simulations was set to half the actual field in the calculations presented here.

Simulations were performed for sextupole only, sextupole+solenoid, and sextupole+solenoid+sextupole configurations. It was found that the sextupole skew angle has minor effect on the sextupole emittance growth. This is because the quadratic radial field generates most of the growth. However, the corrector rotation angle is important when correcting rotated sextupole fields. Simulations show that a rotated corrector sextupole is needed to completely cancel the stray sextupole field emittance. In addition, the strength of the corrector sextupole field depends strongly upon the beam size because of the emittance's third-order dependence on it. Therefore, corrector should then be positioned where the beam is large.

**ACKNOWLEDGEMENTS**

The work is supported by DOE under grant No. DE-AC02-76SF00515.